\documentstyle[12pt,epsf]{article}
 \newcommand{\insertplot}[5]{\begin{figure}
 \hfill\hbox to 0.05in{\vbox to #5in{\vfill
 \inputplot{#1}{#4}{#5}}\hfill}
 \hfill\vspace{-.1in}
 \caption{#2}\label{#3}
 \end{figure}}
 \newcommand{\inputplot}[3]{
 \special{ps: plotfile #1}
}

\begin{document}

\title{STATIC AXIALLY SYMMETRIC SOLUTIONS 
OF EINSTEIN-YANG-MILLS-DILATON THEORY}
\vspace{1.5truecm}
\author{
{\bf Burkhard Kleihaus, and Jutta Kunz}\\
Fachbereich Physik, Universit\"at Oldenburg, Postfach 2503\\
D-26111 Oldenburg, Germany}

\vspace{1.5truecm}


\maketitle
\vspace{1.0truecm}

\begin{abstract}
We construct static axially symmetric solutions of
SU(2) Einstein-Yang-Mills-dilaton theory.
Like their spherically symmetric counterparts,
these solutions are nonsingular and asymptotically flat.
The solutions are characterized by the winding number $n$
and the node number $k$ of the gauge field functions.                          
For fixed $n$ with increasing $k$ the solutions tend
to ``extremal'' Einstein-Maxwell-dilaton black holes
with $n$ units of magnetic charge.
\end{abstract}
\vfill
\noindent {Preprint hep-th/9612101} \hfill\break
\vfill\eject

\section{Introduction}

SU(2) Einstein-Yang-Mills (EYM) theory possesses 
regular static spherically symmetric solutions \cite{bm}.
These solutions are aymptotically flat
and have non-trivial magnetic gauge field configurations,
but no global charge.
They have a high-density interior region,
followed by a near-field region with approximately
Reissner-Nordstr\o m metric and a far-field region
with approximately Schwarzschild metric \cite{bm}.

To every regular solution in SU(2) EYM theory,
there exists a corresponding family of black hole solutions
with regular event horizon $x_{\rm H}>0$ \cite{su2}.
Outside their event horizon
these black hole solutions possess non-abelian hair.
Like the regular solutions,
the black hole solutions are unstable \cite{strau}.

Like EYM theory, Einstein-Yang-Mills-dilaton (EYMD)
possesses static, spherically symmetric
non-abelian regular and black hole solutions \cite{eymd}.
Here the dilaton coupling constant $\gamma$ represents a parameter.
In the limit $\gamma \rightarrow 0$ 
the dilaton decouples and EYM theory is obtained,
for $\gamma = 1$ contact with the low energy effective action
of string theory is made,
and in the limit $\gamma \rightarrow \infty$ gravity decouples and
Yang-Mills-dilaton (YMD) theory is obtained.

Recently we have shown, that YMD theory possesses also static
axially symmetric regular solutions \cite{kkd}. These solutions
are labelled by the winding number $n>1$ and the node number $k$
of the gauge fields. For $n=1$ one obtains the static
spherically symmetric solutions of YMD theory \cite{ymd}.
The axially symmetric solutions have a torus-like shape.
Choosing the $z$-axis as the symmetry axis,
the energy density has a strong peak along the $\rho$-axis
and decreases monotonically along the $z$-axis.

The existence of the regular static axially symmetric YMD solutions
strongly suggests the existence of the corresponding
EYMD solutions for finite values of the dilaton coupling constant.
In this letter we present strong (numerical) evidence,
that such regular gravitating axially symmetric solutions
indeed exist in EYMD theory and also in EYM theory.

\section{\bf Axially symmetric ansatz}

We consider the SU(2) Einstein-Yang-Mills-dilaton action
\begin{equation}
S=\int \left ( \frac{R}{16\pi G} + L_M \right ) \sqrt{-g} d^4x
\   \end{equation}
with
\begin{equation}
L_M=-\frac{1}{2}\partial_\mu \Phi \partial^\mu \Phi
 -e^{2 \kappa \Phi }\frac{1}{2} {\rm Tr} (F_{\mu\nu} F^{\mu\nu})
\ . \end{equation}

To obtain static axially symmetric solutions we employ isotropic
coordinates and adopt the metric
\begin{equation}
ds^2=
  - f dt^2 +  \frac{m}{f} \left( d \rho^2+ dz^2 \right)
           +  \frac{l}{f} \rho^2 d\phi^2
\ , \label{metric} \end{equation}
with $f$, $m$ and $l$ being only functions of $\rho$ and $z$.
The corresponding ansatz for the purely magnetic
gauge field ($A_0=0$) is \cite{rr,kkd}
\begin{equation}
A_\rho= \frac{1}{2}\tau^n_\phi w^3_1
\ , \end{equation}
\begin{equation}
A_z= \frac{1}{2}\tau^n_\phi w^3_2
\ , \end{equation}
\begin{equation}
A_\phi= \frac{1}{2}\tau^n_\rho \rho w^1_3
       +\frac{1}{2}\tau_z      \rho w^2_3
\ , \end{equation}
with the Pauli matrices $\vec \tau$ and
$\tau^n_\rho = \vec \tau \cdot ( \cos n \phi, \sin n \phi,0)$,
$\tau^n_\phi = \vec \tau \cdot (-\sin n \phi, \cos n \phi,0)$.
The four functions $w^i_j$ and the dilaton function $\Phi$
depend only on $\rho$ and $z$.

With this ansatz the energy density $\epsilon =-T_0^0=-L_M$ becomes
\begin{eqnarray}
-T_0^0& = & \frac{f}{2m} \left[
 (\partial_\rho \Phi )^2 + (\partial_z \Phi )^2 \right]
       + e^{2 \kappa \Phi} \frac{f^2}{2 m^2}
 (\partial_\rho w_2^3 - \partial_z w_1^3 )^2 
\nonumber \\
      & + &e^{2 \kappa \Phi} \frac{f^2}{2 m l} \left [
  \left( \partial_\rho w_3^1 + \frac{( n w_1^3 + w_3^1 )}{\rho}
	 - g w_1^3 w_3^2 \right)^2 
+ \left (\partial_\rho w_3^2 + \frac{            w_3^2  }{\rho}
	 + g w_1^3 w_3^1 \right)^2  \right.
\nonumber \\
      & + & \left.
  \left (\partial_z    w_3^1 + \frac{ n w_2^3           }{\rho}
	 - g w_2^3 w_3^2 \right)^2 
+ \left (\partial_z    w_3^2 
	 + g w_2^3 w_3^1 \right)^2
      \right]
\ . \end{eqnarray}

The system possesses a residual abelian gauge invariance 
\cite{kkb,kk,kkd}.
With respect to the transformation
\begin{equation}
 U= e^{i\Gamma(\rho,z) \tau^n_\phi}
\   \end{equation}
the functions
$(\rho w_3^1,\rho w_3^2-n/g)$ transform like a scalar doublet,
and the functions $(w_1^3,w_2^3)$ transform
like a 2-dimensional gauge field.
We fix the gauge by choosing the
gauge condition \cite{kkb,kk,kkd}
\begin{equation}
\partial_\rho w_1^3 + \partial_z w_2^3 =0 
\ . \end{equation}

To make contact with the spherically symmetric case $n=1$,
we introduce the
coordinates $r$ and $\theta$ ($\rho=r \sin \theta$, $z=r \cos \theta$)
and the gauge field functions $F_i(r,\theta)$ \cite{kk,kkd}
\begin{eqnarray}
w_1^3 \  & 
= &  \ \ {1 \over{gr}}(1 - F_1) \cos \theta \ , \ \ \ \ 
w_2^3 \    
= - {1 \over{gr}} (1 - F_2)\sin \theta    \ ,     
\nonumber \\
w_3^1 \  & 
= & - {{ n}\over{gr}}(1 - F_3)\cos \theta    \ , \ \ \ \
w_3^2 \    
=  \ \ {{ n}\over{gr}}(1 - F_4)\sin \theta 
\ . \end{eqnarray}
The spherically symmetric ansatz of ref.~\cite{eymd} is recovered
for $F_1(r,\theta)=F_2(r,\theta)=F_3(r,\theta)=F_4(r,\theta)=w(r)$,
$\Phi(r,\theta)=\phi(r)$ and $n=1$.

The above ansatz and gauge choice
yield a set of coupled partial differential equations
for the metric and the matter field functions.
To obtain regular asymptotically flat solutions
with finite energy density
we impose at the origin ($r=0$) the boundary conditions
\begin{equation}
\partial_r f= \partial_r m= \partial_r l= \partial_r \Phi =0 \ , \ \ \
F_1=F_2=F_3=F_4=1
\ , \end{equation}
and at infinity ($r=\infty$)
\begin{equation}
f=m=l=1 \ , \ \ \ \Phi=0 \ , \ \ \
F_1=F_2=F_3=F_4=\pm 1
\ , \end{equation}
further we impose for all functions that their derivatives
with respect to $\theta$ vanish on the $\rho$- and the $z$-axis
\cite{kkd}.
The boundary conditions for the gauge field functions at infinity
imply, that the solutions are magnetically neutral.
Note, that a finite value of the dilaton field at infinity
can always be transformed to zero via
$\Phi \rightarrow \Phi - \Phi(\infty)$, 
$r \rightarrow r e^{-\kappa \Phi(\infty)} $.

The mass $M$ of the regular axially symmetric
solutions can be obtained directly from
the total energy-momentum ``tensor'' $\tau^{\mu\nu}$
of matter and gravitation,
$M=\int \tau^{00} d^3r$ \cite{wein},
or from
$
M = - \int \left( 2 T_0^{\ 0} - T_\mu^{\ \mu} \right)
   \sqrt{-g} dr d\theta d\phi
.$
Both expressions give the same values for the mass
of the solutions.

We now remove the dependence on the coupling constants
$\kappa$ and $g$ from the differential equations
by changing to the dimensionless coordinate 
$x=(g/\sqrt{4\pi G}) r$, 
the dimensionless dilaton function $\varphi = \sqrt{4\pi G} \Phi$ and  
the dimensionless coupling constant
$\gamma =\kappa/\sqrt{4\pi G}$
($\gamma=1$ corresponds to string theory).
The dimensionless mass is then given by $\mu =(g/\sqrt{4\pi G}) G M$.

In the spherically symmetric case the following relations 
between the metric and the dilaton field hold \cite{kks3}
\begin{equation}
\varphi(x) = \frac{1}{2} \gamma \ln(-g_{tt})
\ , \label{res2} \end{equation}
\begin{equation}
\mu = \frac{1}{2} x^2 \partial_x f |_\infty
    = \frac{1}{\gamma} x^2 \partial_x \varphi |_\infty
    = \frac{D}{\gamma}
\ , \label{res1} \end{equation}
where $D$ is the dilaton charge.
These relations also hold for the regular axially symmetric solutions
considered here.
Their derivation is based on the equation of motion
of the dilaton field and will be given elsewhere \cite{kknew}.

\section{\bf Solutions}

Subject to the above boundary conditions,
we solve the equations numerically.
To map spatial infinity to the finite value $\bar{x}=1$,
we employ the radial coordinate $\bar{x} = \frac{x}{1+x}$.
The numerical calculations are based on the Newton-Raphson
method. The equations are discretized on a non-equidistant
grid in $\bar{x}$ and  $\theta$.
Typical grids used have sizes $150 \times 30$, 
covering the integration region 
$0\leq\bar{x}\leq 1$ and $0\leq\theta\leq\pi/2$.
The numerical error for the functions is estimated to be 
on the order of $10^{-3}$ and $10^{-2}$ for $k<4$ and $k=4$,
respectively. 

\begin{figure}
\centering
\epsfysize=11cm
\mbox{\epsffile{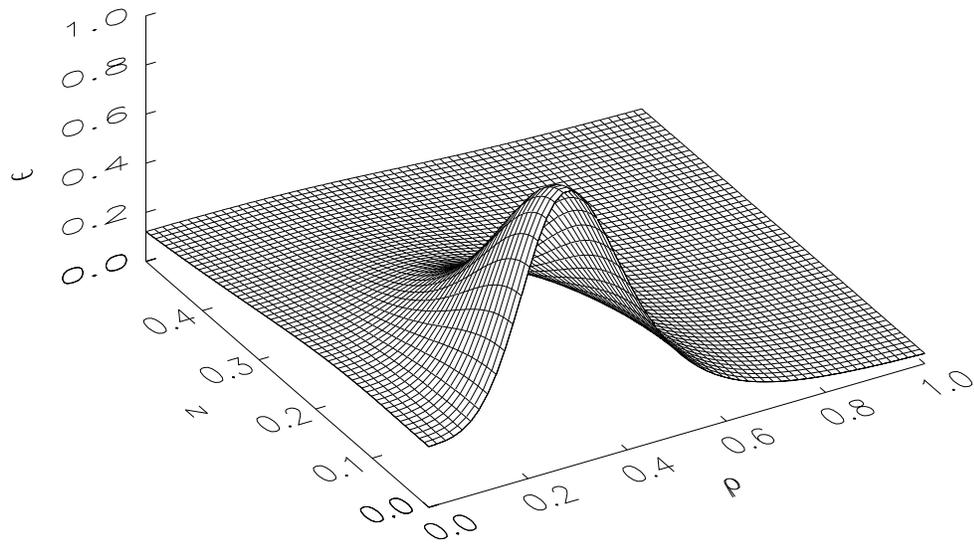}}
\caption{\label{eps}
The energy density $\epsilon=-T_0^0$
is shown as a function of the dimensionless coordinates
$\rho$ and $z$ for the EYMD solution with
winding number $n=3$, node number $k=3$ and $\gamma=1$.
}
\end{figure}

In Tables~1 and 2 we show the dimensionless mass 
of a subset of the regular axially symmetric
solutions obtained so far.
The energy density $\epsilon=-T_0^0$ of the solutions
has a strong peak along the $\rho$-axis,
and it decreases monotonically along the $z$-axis.
Thus equal density contours reveal a torus-like shape
of the solutions.
As a typical example we show the energy density $\epsilon$
of the solution with $n=3$, $k=3$
and dilaton coupling constant $\gamma=1$ in Fig.~\ref{eps}.

With $n$ and $\gamma$ fixed and increasing $k$, the location of the
peak of the energy density moves inward and the peak
increases in height, whereas with fixed $k$ and $\gamma$
and increasing $n$ the peak of the energy density moves outward
and decreases in height.
This is demonstrated in Table~2.

The gauge field functions $F_i$ and the dilaton function $\varphi$
of the regular axially symmetric EYMD and EYM solutions 
look similar to those of the corresponding YMD solutions \cite{kkd}.
Like the dilaton function $\varphi$, the metric functions
do not exhibit a strong angular dependence.
These functions will be exhibited elsewhere \cite{kknew}.

\begin{figure}
\centering
\epsfysize=11cm
\mbox{\epsffile{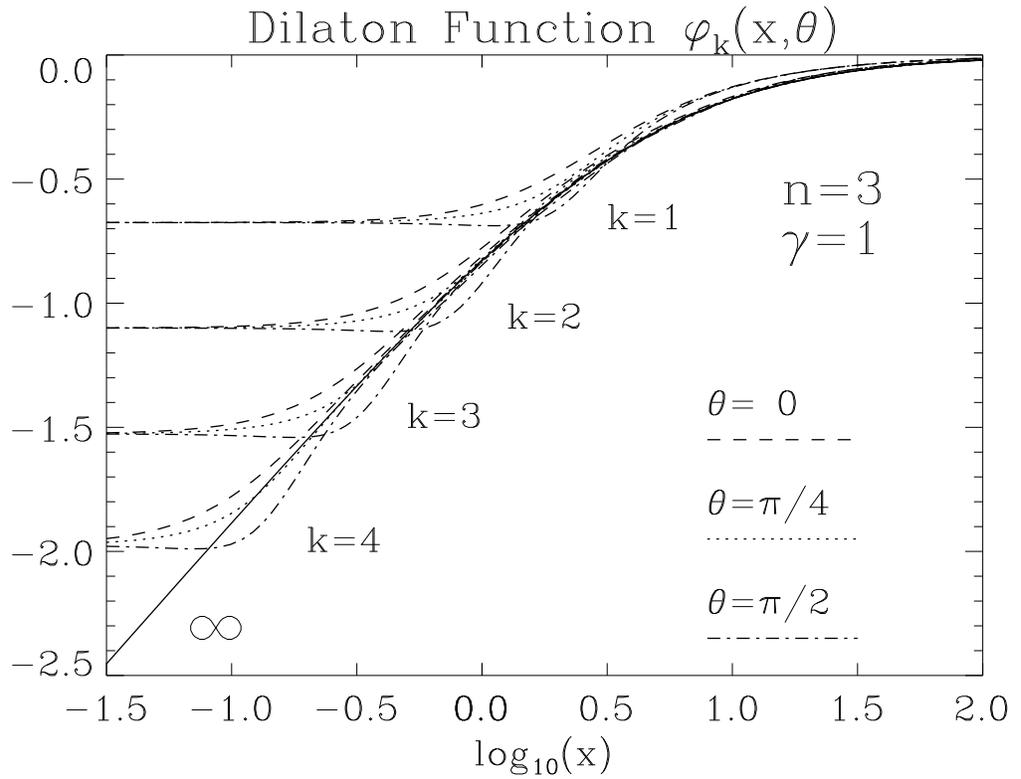}}
\caption{\label{phi}
The dilaton functions $\varphi_k(x,\theta)$ for the EYMD solutions with
winding number $n=3$, node numbers $k=1-4$ and $\gamma=1$
are shown as a function of the dimensionless coordinate $x$.
The dashed, the dotted and the dash-dotted lines represent
the angles $\theta=0$, $\theta= \pi/4$ and $\theta= \pi/2$,
respectively. Also shown is the limiting function
$\varphi_\infty(x)$ (solid line).
}
\end{figure}

For fixed $n$ and $\gamma$, with increasing $k$
the sequence of axially symmetric solutions
tends to a limiting solution.
The gauge field functions $(F_i)_k$ tend to the 
limiting function $F_\infty=0$.
(Because of the boundary conditions at the origin and at infinity,
they approach the limiting function nonuniformly.)
The dilaton functions $\varphi_k$ tend to the
limiting function $\varphi_\infty$,
which represents the dilaton function
of the ``extremal'' EMD solution \cite{emd}
with $n$ units of magnetic charge and the same value of $\gamma$.
This is demonstrated in Fig.~\ref{phi} for $n=3$ and $\gamma=1$.
We observe, that $\varphi_k$ deviates from $\varphi_\infty$
only in an inner region, which decreases exponentially with $k$.

For finite values of $\gamma$ and fixed $n$,
the sequences of solutions thus approach as limiting solutions
the ``extremal'' EMD solutions \cite{emd} with magnetic charge $n$
and the same $\gamma$. This generalizes the corresponding
observation for the spherically symmetric solutions with $n=1$
\cite{eymd,kks3}.
For $\gamma=0$ the limiting solutions represent 
the exterior of extremal Reissner-Nordstr\o m
black holes with charge $n$.
For the YMD solutions the limiting solutions are magnetic
monopoles with $n$ units of charge \cite{kkd}.

The limiting values for the mass, $\mu=n/\sqrt{1+\gamma^2}$ \cite{emd},
represent upper bounds for the sequences,
as observed from Tables~1 and 2.
The larger $n$, the slower is the convergence to the limiting solution.
Further details on the convergence properties
of the sequences of solutions will be given elsewhere \cite{kknew}.

\section{Conclusions}

In addition to the known static spherically symmetric solutions,
both EYMD and EYM theory possess
sequences of regular static axially symmetric solutions.
These sequences are characterized by the winding number
$n>1$, and the solutions within each sequence by the node number $k$.
(For $n=1$ the spherically symmetric solutions are recovered.)
For fixed $n$ and $\gamma$,
with increasing $k$ the solutions tend to the
``extremal'' Einstein-Maxwell-Dilaton solution \cite{emd}
with $n$ units of magnetic charge and the same $\gamma$.

The multisphalerons have a torus-like shape. Apart from that,
many properties of the axially symmetric solutions are similar
to those of their spherically symmetric counterparts.
In particular,
there is all reason to believe, that these regular static
axially symmetric EYMD and EYM solutions are unstable.
Since we can also associate the Chern-Simons number 
$N_{\rm CS}=n/2$ \cite{kk} (for odd $k$ \cite{gv})
with these solutions, we interpret them 
as {\sl gravitating multisphalerons}.

Having constructed the axially symmetric solutions
in EYMD and EYM theory, it appears straightforward
to construct analogous solutions is theories
with a Higgs field. We therefore expect to find
gravitating axially symmetric multimonopoles
(for the case of a Higgs triplet) or
gravitating axially symmetric multisphalerons
(for the case of a Higgs doublet).
Similarly there should be
gravitating axially symmetric multiskyrmions.
Work along these lines is in progress.

We consider the above set of solutions to be the simplest type of
gravitating non-spherical regular solutions of EYMD and EYM theory.
We conjecture, that there are gravitating regular solutions
with much more complex shapes and only discrete symmetries left.
This conjecture is based on the observation,
that for some types of solitons in flat space
the symmetry structure of the (energetically lowest) solutions
becomes increasingly complex with increasing 
winding number or charge $n$.
For instance for skyrmions,
the lowest $n=1$ solution is spherically
symmetric, the lowest $n=2$ solution has axial symmetry,
and the lowest $n \ge 3$ solutions respect only
discrete crystal-like symmetries \cite{bc}.

But EYMD and EYM theory also possess black hole solutions.
The non-abelian spherically symmetric black hole solutions
may be regarded as black holes inside sphale\-rons \cite{gv}.
We conjecture, that also the gravitating axially symmetric solutions
can accommodate black holes in their interior.
And this conjecture naturally extends to gravitating solutions
with more complex shapes and less symmetry.
The existence of such black hole solutions without rotational 
symmetry inside multimonopoles has also been
conjectured from a stability argument \cite{ewein}.

\newpage
\begin{table}[p!]
\begin{center}
\begin{tabular}{|c|c|ccc|c|} \hline
\multicolumn{1}{|c|} { $ $ }&
\multicolumn{1}{ c|} { $EYM $ }&
\multicolumn{3}{ c|}  {$EYMD$ } & 
\multicolumn{1}{ c|} { $YMD$ }\\   
 \hline
 $k/\gamma$ &  $0$     &    $0.5$  &  $1.0$   & $2.0$ &   $\infty$ \\
 \hline
 $1$        &  $1.870$ &   $1.659$ &  $1.297$ & $0.811$ & $1.800$ \\ 
 $2$        &  $2.524$ &   $2.250$ &  $1.770$ & $1.114$ & $2.482$ \\  
 $3$        &  $2.805$ &   $2.505$ &  $1.976$ & $1.247$ & $2.785$ \\  
 $4$        &  $2.922$ &   $2.611$ &  $2.063$ & $1.304$ & $2.913$ \\   
 \hline  
 $-$        &  $3    $ &   $2.683$ &  $2.121$ & $1.342$ & $3    $ \\ 
\hline
\end{tabular}
\end{center} 
\vspace{1.cm} 
{\bf Table 1}\\
The dimensionless mass $\mu$ of the EYMD solutions
with winding number $n=3$ and up to 4 nodes
for several values of the dilaton coupling constant $\gamma$.
For comparison, the last row gives the mass of the limiting
solutions, the first column gives the mass
of the EYM solutions, and the last column the
scaled mass of the corresponding YMD solutions \cite{kkd}.
\end{table}
\begin{table}[p!]
\begin{center}
\begin{tabular}{|c|cccc|} 
 \hline 
\multicolumn{1}{|c|} { $ $ }&
\multicolumn{4}{ c|}  {$\mu$ } \\
 \hline 
$k/n$ &  $1$  & $2$ & $3$ & $4$ \\
 \hline 
$1$ &  $0.577$   & $0.961$  & $1.297$  & $1.607$  \\ 
$2$ &  $0.685$   & $1.262$  & $1.770$  & $2.239$  \\ 
$3$ &  $0.703$   & $1.365$  & $1.976$  & $2.549$  \\  
$4$ &  $0.707$   & $1.399$  & $2.063$  & $2.698$  \\ 
\hline  
$-$ &  $0.707$   & $1.414$  & $2.121$  & $2.828$  \\
\hline
\multicolumn{1}{|c|} { $ $ }&
\multicolumn{4}{ c|}  {$\epsilon_{max}(\rho_{max})$ } \\
 \hline 
$k/n$ &  $1$  & $2$ & $3$ & $4$ \\
\hline
$1$ &  $1.075 \ (0.)$   & $0.177 \ (0.90)$  & $0.098 \ (1.59)$ 
 & $0.072 \ (2.37)$  \\ 
$2$ &  $11.63 \ (0.)$   & $0.910 \ (0.30)$  & $0.380 \ (0.66)$  
& $0.235 \ (1.10)$  \\ 
$3$ &  $79.70 \ (0.)$   & $3.443 \ (0.09)$  & $1.124 \ (0.28)$  
& $0.601 \ (0.51)$  \\  
$4$ &  $498.2 \ (0.)$   & $12.01 \ (0.03)$  & $3.064 \ (0.11)$  
& $1.435 \ (0.24)$  \\ 
\hline
\end{tabular}
\end{center} 
{\bf Table 2}\\
The dimensionless mass $\mu$, 
 the maximum of the energy density $\epsilon_{\rm max}$
and the location $\rho_{max}$ of the maximum of the energy density 
of the EYMD solutions of the sequences $n=1-4$
with node numbers $k=1-4$ and $\gamma=1$.
For comparison, the mass of the limiting
solutions is also shown.
Note, that $\mu/n$ decreases with $n$ for fixed finite $k$.
\end{table}

\clearpage
\newpage

\end{document}